\documentclass{aa}
\usepackage{graphicx}
\usepackage{txfonts}
\usepackage{natbib}
\newcommand{\eqb}{\begin{eqnarray}}
\newcommand{\eqe}{\end{eqnarray}}
\newcommand{\melec}{m_{\rm e}}
\newcommand{\mprot}{m_{\rm p}}
\newcommand{\gammap}{\gamma_{\rm p}}
\newcommand{\gammae}{\gamma_{\rm e}}
\newcommand{\nelec}{n_{\rm e}}
\newcommand{\nprot}{n_{\rm p}}
\newcommand{\nprin}{n_{\rm p,0}}
\newcommand{\nprincrit}{n_{\rm p,0}^{\rm crit}}
\newcommand{\nphot}{n_\gamma}
\newcommand{\teesc}{t_{\rm e,esc}}
\newcommand{\tpesc}{t_{\rm p,esc}}
\newcommand{\sigmaT}{\sigma_{\rm T}}

\newcommand{\protin}{Q_{\rm {p,0}}}
\newcommand{\elecrat}{Q_{\rm e}^{\rm ext}}
\newcommand{\elecin}{Q_{\rm {e,0}}}
\newcommand{\protrat}{Q_{\rm p}}
\newcommand{\tempbb}{T_{\rm BB}}
\newcommand{\lbb}{\ell_{\rm BB}}
\newcommand{\lelecext}{\ell_{\rm e}^{\rm ext}}
\newcommand{\lpairinj}{\ell_{\rm e}^{\rm BH}}
\newcommand{\lprot}{\ell_{\rm p}}
\newcommand{\lphot}{\ell_{\gamma}}
\newcommand{\protdens}{n_{\rm p}}
\newcommand{\photdens}{n_{\rm \gamma}}
\newcommand{\gcrit}{\gamma_{\rm crit}}
\newcommand{\bcrit}{B_{\rm cr}}
\newcommand{\qbhinj}{Q_{\rm e}^{\rm BH}}
\newcommand{\protloss}{L_{\rm p}^{\rm BH}}
\newcommand{\qsyninj}{Q_{\gamma}^{\rm syn}}
\newcommand{\lsynlos}{L_{\rm e}^{\rm syn}}
\newcommand{\qicsinj}{Q_{\gamma}^{\rm ics}}
\newcommand{\licslos}{L_{\rm e}^{\rm ics}}
\newcommand{\qgginj}{Q_{\rm e}^{\gamma\gamma}}
\newcommand{\lgglos}{L_{\gamma}^{\gamma\gamma}}
\newcommand{\qannih}{Q_{\gamma}^{\rm ann}}
\newcommand{\lannih}{L_{\rm e}^{\rm ann}}
\newcommand{\lssa}{L_{\gamma}^{\rm ssa}}
\newcommand{\tthom}{\tau_{\rm T}}
\newcommand{\lb}{\ell_{\rm B}}

\newcommand{\gammapmin}{\gamma_{\rm p,min}}
\newcommand{\gammapmax}{\gamma_{\rm p,max}}
\newcommand{\gammaemin}{\gamma_{\rm e,min}}
\newcommand{\gammaemax}{\gamma_{\rm e,max}}

\newcommand{\xmax}{x_{\rm max}}

\newcommand{\tcross}{t_{\rm cr}}

\newcommand{\tcool}{t_{\rm cool}}

\newcommand{\myemail}{amastich@phys.uoa.gr}

\begin{document}
\title{Spectral and temporal signatures
of ultrarelativistic protons in compact sources} 
\subtitle{I. Effects of
Bethe-Heitler pair production}

\author{A. Mastichiadis\inst{1}
\and
R.J. Protheroe\inst{2}
\and
J.G. Kirk\inst{3}}
\offprints{A. Mastichiadis}
\institute{Department of Physics, University of Athens, Panepistimiopolis,
  GR 15783, Zografos, Greece\\
\email{\myemail}
\and
Department of Physics and Mathematical Physics, University of Adelaide,
North Terrace, Adelaide, SA 5005, Australia
\and
Max-Planck-Institut f\"ur Kernphysik, Postfach 103980,
69029 Heidelberg, Germany}
\date{Received ?; accepted ?}
\titlerunning{Hadrons}
\authorrunning{Mastichiadis et al.}

\abstract{
We present calculations of the spectral and temporal radiative 
signatures expected from ultrarelativistic protons in compact sources. 
The coupling between 
the protons
and the leptonic component is assumed to occur via Bethe-Heitler 
pair production. This process is treated by modeling the results of 
Monte-Carlo simulations and incorporating them in a time-dependent
kinetic equation, that we subsequently solve numerically. 
Thus, the present work is, in many
respects, an extension of the leptonic `one-zone' models to include 
hadrons. Several examples of 
astrophysical importance are presented, such as the signature
resulting from the cooling of relativistic protons
on an external black-body field and that of their cooling in the presence of 
radiation from 
injected electrons. We also investigate and refine the threshold conditions
for the
'Pair Production/Synchrotron' feedback loop 
which operates when relativistic protons 
cool efficiently on the synchrotron radiation of the internally produced
Bethe-Heitler pairs. We demonstrate that an additional component of injected 
electrons lowers the threshold for this instability. 
}
\maketitle

\section{Introduction}

The spectral energy distribution (SED) of powerful AGN such as
flat-spectrum radio quasars and blazars has a double humped
appearance with the low energy part extending from the radio to
UV (or in extreme cases to X-rays), and a high energy part
extending from X-rays to $\gamma$-rays.  In AGN with relativistic 
jets closely aligned to the line of sight the emission is dominated by 
non-thermal
radiation, with the low energy hump being mainly synchrotron  
radiation.  If the alignment is not so close, a thermal component of
UV radiation from an accretion disk may dominate.  The
non-thermal components can be strongly variable, probably
originating in the jet.  

These observations indicate that the jets of blazars
act as efficient particle accelerators. Furthermore
the gamma-ray observations
in the GeV \citep{hartmanetal99}
and TeV regime \citep{horanweekes04} imply that the accelerated
particles can reach very high energies. Models
involving electron radiation  
can adequately explain both this
high energy emission and the coordinated multiwavelength campaigns
\citep{mastichiadiskirk97,tavecchioetal01,krawczynskietal02}. 
The usual assumption is then that the high energy
part of the SED is due to inverse Compton scattering of the low energy
part of the SED (synchrotron self-Compton model) possibly supplemented
by inverse Compton scattering of external photons (external Compton 
models)
for example from the disk, either directly or scattered by clouds.
Despite these successes,
the question of the role of a possible relativistic hadronic
component remains an open one.

In principle, sites of electron acceleration 
may
accelerate 
protons as well. 
Consequently,
models in which the high energy part, and some fraction of the
low energy part, of the SED is due to acceleration and
interaction of protons in the jet have also been proposed. Some of
these models invoke interactions with ambient matter 
\citep{beallbednarek99,pohlschlickeiser00,schusteretal02} 
but they require high mass densities in the
jet to be viable.  Here we concentrate on hadronic models in
which the protons interact with low energy photons via
Bethe-Heitler pair production.

As with leptonic models, the target photons may
be produced inside the emission region in the jet 
or may originate from outside of the jet, e.g., from an accretion disk 
\citep{protheroe97,bednarekprotheroe99,
atoyandermer01,neronovsemikoz02}.  
For internally produced target
photons, synchrotron emission by a co-accelerated population of
electrons is assumed \citep{mannheimbiermann92,mannheim93,mannheim95}.  The high energy hump of the SED then results from
electromagnetic cascading of gamma-rays from $\pi^0$ decay and
electrons from Bethe-Heitler pair production and $\pi^\pm \to
\mu^\pm \to e^\pm$ decay in the radiation and magnetic field of
the blob.  Neutrinos and cosmic rays would also be emitted as a
result of the neutrinos from the $\pi^\pm$ and $\mu^\pm$ decays,
and neutrons produced in $p\gamma \to n \pi^+$ interactions if
the threshold for pion photoproduction is exceeded 
\citep{eichlerwiita78,
sikoraetal87,
sikoraetal89,kirkmastichiadis89,
begelmanetal90,giovanonikazanas90,protheroeszabo92,szaboprotheroe94,
waxmanbahcall99,
mannheimetal01,atoyandermer03,protheroe04}.  
For protons to be accelerated to energies sufficient
to exceed the Bethe-Heitler and
photo-pion-production thresholds, relatively high magnetic fields
are required. In proton synchrotron blazar models the magnetic
field is sufficiently high such that the the high energy part of
the SED has a major contribution also due to synchrotron
radiation by protons 
\citep{mueckeprotheroe00,mueckeprotheroe01,aharonian00,
reimeretal03}.  
All of the work described
above assumes the emission has reached a steady state, 
and
that
the target photon fields are steady.  In reality, 
the strong variability displayed by these sources mandates a time-dependent
calculation, that, ideally,
should be done
self-consistently, with internally produced radiation fields
contributing alongside external ones to the target radiation field.

Time-dependent codes that solve the kinetic equations   
describing electrons and photons and their interactions 
have been developed and successfully applied to AGN 
\citep{mastichiadiskirk97,krawczynskietal02}.
 However, 
codes of this type that also account for hadronic interactions
have been neglected.
One reason for this is that whereas the modeling of 
leptonic processes is relatively straightforward
\citep[e.g.,][]{lightmanzdziarski87,coppiblandford90},
photo-hadronic and hadron-hadron
interactions are much more complex. To date, all attempts have
used approximations of uncertain accuracy \citep[e.g.,][]{sternsvensson91}
but the use of 
Monte-Carlo event generators which model in 
detail electromagnetic \citep{szaboprotheroe94,protheroejohnson96} and
hadronic interactions 
\citep{mueckeetal00}
opens up the possibility of 
extracting accurate descriptions of the fundamental interactions suitable
for incorporation into a kinetic code.

Motivated by these developments we investigate the consequences 
of the presence of relativistic hadrons in compact sources by
incorporating new results from Monte-Carlo simulations
into a time-dependent
code which follows the evolution of relativistic
hadrons, electrons and photons by solving the
appropriate kinetic equations.
In the present paper we investigate as a first step, the case in which the only 
channel of coupling between hadrons and leptons is the Bethe-Heitler 
pair-production process, leaving the investigation of photo-meson production 
for a future paper. Although this is not a complete description 
of hadronic models it nevertheless enables one to draw useful and 
interesting new results.

The present paper is structured as follows: 
In Section~2 we present the
numerical code that solves simultaneously in a self-consistent 
manner the coupled, time-dependent kinetic equations for each species, 
i.e. protons, electrons and photons.
In Section~3 we present the Monte-Carlo results for the 
Bethe-Heitler  process and show how these can be incorporated in the
kinetic equations. In Section~4 we present 
some results for the case in which relativistic 
protons interact with an external black-body radiation field.
In Section~5, we present a numerical analysis of the 
'Pair-Production/Synchrotron' instability. The case of simultaneous
injection of relativistic protons and electrons is examined in
Section~6 and the main conclusions are summarized in Section~7.

\section{The kinetic equations for electrons, protons and photons}

The kinetic equations describing a homogeneous source
region containing protons, electrons and photons were
formulated and solved numerically by 
\citet[henceforth MK95]{mastichiadiskirk95}.
\defcitealias{mastichiadiskirk95}{MK95}
We follow the same method, using an improved description
of the microscopic processes. The equations to be solved can
be written in the generic form
\eqb
{{\partial n_i}\over{\partial t}} + L_i +Q_i=0
\eqe
where the index $i$ can be any one of the 
subscripts `p', `e' or `$\gamma$' referring to protons,
electrons or photons respectively.
The operators $L_i$  denote losses and escape from the system
while $Q_i$ denote injection and source terms. These are
defined below.

The unknown functions $n_i$ are the differential
number densities of the three species, normalised as follows:
\eqb
\noalign{\hbox{Protons:}}
&\nprot ^*(\gammap,t)d\gammap\,=\,\sigmaT
R\nprot(
E_{\rm p},t) dE_{\rm p} 
&\textrm{with }
\gammap\,=\,{{E_{\rm p}}\over{\mprot c^2}}
\\
\noalign{\hbox{Electrons:}}
&\nelec^*(\gammae,t)d\gammae=\sigmaT R
\nelec
(E_{\rm e},t) dE_{\rm e}&
\textrm{with } \gammae\,=\,{{E_{\rm e}}\over{\melec c^2}}
\\
\noalign{\hbox{Photons:}}
&\nphot^*(x,t)dx\,=\,\sigmaT R\nphot(\epsilon_\gamma,t)
d\epsilon_\gamma&
\textrm{with } x\,=\,{{\epsilon_\gamma}\over{\melec c^2}}
\eqe
and the time $t$ has been normalised in all equations to the light-crossing
time of the source $\tcross=R/c$.

The physical processes to be included in the kinetic equations are:
\begin{enumerate}
\item
Proton-photon (Bethe-Heitler) pair production which acts as a loss term
for the protons ($\protloss$) and an injection term for the electrons 
($\qbhinj$) 
\item
Synchrotron radiation which acts as an energy loss term for electrons ($\lsynlos$)
and as a source term for photons ($\qsyninj$)
\item
Synchrotron self absorption which acts as an absorption term for photons ($\lssa$)
\item
Inverse Compton scattering (in both the Thomson and Klein-Nishina regimes)
which acts as an energy loss term for electrons ($\licslos$)
and as a source term for high energy photons
and a loss term for low energy photons, both effects
included in  $\qicsinj$
\item
Photon-photon pair production
which acts as an injection term for electrons ($\qgginj$)
and as an absorption term for photons ($\lgglos$)
\item
Electron-positron annihilation 
which acts as a sink term for electrons ($\lannih$)
and as a source term for photons ($\qannih$)
\item
Compton scattering of radiation on the cool pairs, which impede 
the free escape of photons
from the system. This effect is treated approximately
by multiplying the photon escape term
by the factor $(1+H(1-x)\tthom/3)^{-1}$, where $\tthom$ the
Thomson optical depth, while $H(1-x)$ is the step-function.
\citep{lightmanzdziarski87}
\end{enumerate}

With the inclusion of the above terms the kinetic equations become, 
for each species
(from now on we refer only to normalised quantities 
and, for convenience, drop the asterisks)

\begin{itemize}
\item
Protons
\eqb
{{\partial\nprot}\over{\partial t}} + \protloss + {\nprot\over\tpesc}=\protrat
\label{protkinet}
\eqe
\item
Electrons
\eqb
{{\partial\nelec}\over{\partial t}} + \lsynlos + \licslos + \lannih +
{\nelec\over\teesc}=\elecrat + \qbhinj + \qgginj
\label{eleckinet}
\eqe
\item
Photons
\eqb
{{\partial\nphot}\over{\partial t}} + {\nphot\over{1+\tthom/3}} +
\lgglos + \lssa =\qsyninj + \qicsinj +\qannih
\label{photkinet}
\eqe
\end{itemize}

We note the following regarding the above equations
\begin{enumerate}
\item
When the various terms above are written explicitly,
equations (\ref{protkinet}),~(\ref{eleckinet}) and (\ref{photkinet})   
form a non-linear system of coupled integro-differential
equations.
\item 
The various rates conserve the energy  
exchange between the species
-- for example, the amount
of energy lost per second by electrons at each instant
due to synchrotron radiation
is equal to the power
radiated in synchrotron photons, \citepalias[for details see][]{mastichiadiskirk95}
\item
In the absence of the Bethe-Heitler 
pair-production term $\qbhinj$, equations (\ref{eleckinet}) and
(\ref{photkinet}) decouple from the protons (Eq.~\ref{protkinet}) and
the system becomes identical to 
the 'one-zone' time-dependent
leptonic models
\item
Protons are injected via the term $\protrat$ 
with a prescribed distribution in energy. 
Thus, in contrast to 
\citetalias{mastichiadiskirk95}, we do not investigate the effects of particle
acceleration.
\item
Electrons 
may also be injected externally through the
prescribed 
term $\elecrat$. 
However the two other injection terms
in the electron equation ($\qbhinj$ and $\qgginj$) are determined 
self-consistently from the proton and photon distributions.
\item
Both protons and electrons 
(we make no distinction between electrons and positrons
in the present treatment) escape 
from the source region on the timescales $\tpesc$ and $\teesc$ respectively
(given in units of $\tcross$)
\end{enumerate}

We can also define various compactnesses related to the
most important of the above quantities.
So we define the photon compactness as
\citepalias[see][]{mastichiadiskirk95}
\eqb
\lphot&=&{{1}\over{3}}\int dx x {\nphot(x,t)\over
{\tcross(1+H(1-x)\tthom/3)}}
\label{photcompact}
\eqe
while the scaled to electron rest-mass compactness of externally
injected protons is
\eqb
\lprot&=&{\mprot\over3\melec}
\int d\gamma(\gamma-1)\protrat(\gamma).
\label{procompact}
\eqe 
In analogous fashion we can also define the compactness of
the externally injected electrons
\eqb
\lelecext&=&{1\over 3}
\int d\gamma(\gamma-1)\elecrat(\gamma).
\label{extelinj}
\eqe 
Finally in order to calculate the Thomson optical depth 
of the cool pairs we use
\eqb
\tthom=\int_1^{1.26} d\gamma~\nelec(\gamma),
\eqe

Closing this section we note that the treatment of 
synchrotron and inverse Compton scattering has 
been improved over that described by \citetalias{mastichiadiskirk95} in that the full
emissivities are incorporated,
rather than delta-function approximations. However, the main
improvement is in the treatment of the Bethe-Heitler
pair-production process, using Monte-Carlo simulations, as described in the
following section. 

\section { Bethe-Heitler pair production}
\subsection {Monte-Carlo simulations}

\begin{figure}
\centering
\includegraphics[width=8.5cm]{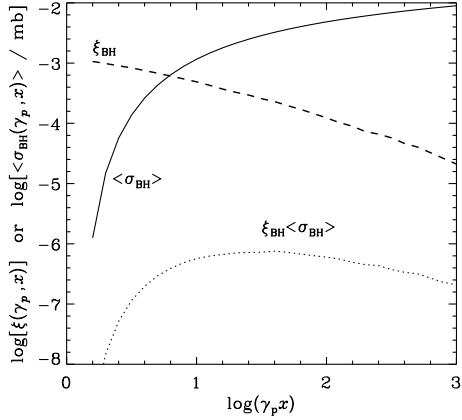}
\caption{The angle-averaged Bethe-Heitler pair-production 
cross section $\langle \sigma_{\rm BH}(\gamma_{\rm p},x)\rangle$ (solid
curve), the angle-averaged Bethe-Heitler pair-production
inelasticity $\xi_{\rm BH}(\gamma_{\rm p},x)$ (dashed curve), and
$\xi_{\rm BH}(\gamma_{\rm p},x)\langle \sigma_{\rm
BH}(\gamma_{\rm p},x)\rangle$ (dotted curve) are plotted as a function
of $\gamma_{\rm p} x$ for $x \ll 1$.}
\label{fig:BHsigma_inel}
\end{figure}

For an isotropic target comprising monoenergetic photons of energy
$\varepsilon=xm_{\rm e}c^2$  the effective cross-section for
interaction of a proton of energy $E=\gamma_{\rm p}m_{\rm p}c^2$ is given by
\begin{equation}
	\langle \sigma_{\rm BH}(\gamma_{\rm p},x)\rangle = {1 \over 2}
\int_{-1}^{\cos\theta_{\rm min}(\gamma_{\rm p},x)} 
(1 - \beta_{\rm p} \cos \theta)\sigma(s)d\cos\theta
\end{equation}
where $\theta$ is the angle between the proton and photon directions, 
$\theta_{\rm min}(\gamma_{\rm p},x)$ is the minimum value of this 
angle consistent with the threshold,  
\begin{equation}
s=m_{\rm p}^2 c^4 + 2 \varepsilon E(1 - \beta_{\rm p} \cos \theta)
\label{eq:s}
\end{equation}
 is the centre of momentum (CM) frame energy squared, $\beta_{\rm p} c$
 is the proton's velocity, and $\sigma_{\rm BH}$ the total cross-section for which we use the Racah formula as parameterized by
 \citet{maximon68} \citep[see Formula 3D-0000 in][]{motzetal69}.  
Changing variables, one obtains the angle-averaged cross-section
\begin{equation}
\langle \sigma_{\rm BH}(\gamma_{\rm p},x)\rangle =  {1 \over 8 \beta_{\rm p} E^2\varepsilon^2}
 \int_{s_{\rm min}}^{s_{\rm max}(\gamma_{\rm p},x)} \sigma(s)(s-m_{\rm p}^2
 c^4)ds, \label{eq:mpl}
\end{equation}
where
\begin{equation}
s_{\rm min}= (m_{\rm p}c^2 + 2 m_{\rm e}c^2)^2 \approx 0.882 \; \rm GeV^2,
\end{equation}
and
\begin{equation}
s_{\rm max}(\gamma_{\rm p},x) = m_{\rm p}^2c^4+2\gamma_{\rm p}m_{\rm p}c^2 xm_{\rm e}c^2 (1+\beta_{\rm p})
\end{equation} 
corresponding to a head-on collision. 
The angle-averaged cross-section is plotted as a function of the product of photon
energy and proton energy in Fig.~\ref{fig:BHsigma_inel}. 
For $x\ll1$, the threshold condition implies $\gamma_{\rm p}\gg1$ and 
$\beta_{\rm p}\approx1$, so that to a first 
approximation, the cross-section is a 
function of $x\gamma_{\rm p}$, rather than 
of $x$ and $\gamma_{\rm p}$ separately. 

Examination of the integrand in Equation \ref{eq:mpl} shows that the 
square of the
total CM frame energy is distributed as
\begin{equation}
	p(s) \propto \sigma(s)(s-m_{\rm p}^2c^4),
\end{equation}
in the range $s_{\rm min}\leq s\leq s_{\rm max}$.  
The Monte Carlo rejection technique is used to sample $s$, and
Equation~\ref{eq:s} is used to find $\theta$.  We then Lorentz transform
the interacting particles to the proton rest frame and sample the positron's
energy from the single-differential cross-section, $d
\sigma / d E_+$, for which we use the Bethe-Heitler formula for an
unscreened point nucleus \citep[Formula 3D-1000 in][]{motzetal69}.
Finally, the positron's direction is sampled from the
double-differential cross-section, $d \sigma / d E_+ d \Omega_+$
for which we use the Sauter-Gluckstern-Hull formula for an
unscreened point nucleus \citep[Formula 3D-2000 in][]{motzetal69}, and
its laboratory frame energy is obtained by a Lorentz
transformation.  For a range of proton energies, the simulation
is repeated a large number of times to build up distributions in
energy of positrons produced in BH pair production.  

\begin{figure}
\centering
\includegraphics[width=8.5cm]{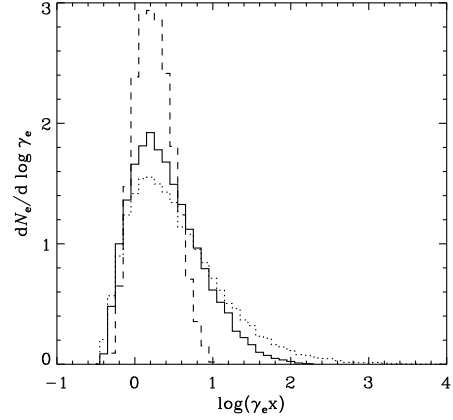}
\caption{The distribution of electron energies is shown for $x=10^{-6}$ and
$\gamma_{\rm p}=10^{6.3}$ (dashed histogram), $10^7$ (solid histogram)
and $10^9$ (dotted histogram).}
\label{fig:BHdist}
\end{figure}

The distribution in energy $\gamma_{\rm e} m_{\rm e}c^2$ of electrons (of
either charge), 
\begin{equation}
f(\gamma_{\rm e}; \gamma_{\rm p},x) \equiv {d N_{\rm e} \over d \gamma_{\rm e}}, 
\label{pairdistr}
\end{equation}
is taken to be twice that for positrons, as discussed in
\citet{protheroejohnson96}, and is plotted in
Fig.~\ref{fig:BHdist} for $x=10^{-6}$ and three $\gamma_{\rm p}$
values.  In a similar calculation using black-body
target photons \citep{protheroejohnson96}, the mean inelasticity was found
to be in excellent agreement with those
calculated analytically
\citep{blumenthal70,chodorowskietal92,rachenbiermann93}.  The mean inelasticity, given by
\begin{equation}
\xi(\gamma_{\rm p},x) = {m_{\rm e} \over m_{\rm p}}
\int d \gamma_{\rm e} {\gamma_{\rm e} \over \gamma_{\rm p}}f(\gamma_{\rm e}; \gamma_{\rm p},x)
\label{inelast}
\end{equation}
is plotted as a function of the product of photon energy and
proton energy in Fig.~\ref{fig:BHsigma_inel} (dashed line) together
with its product (dotted curve) with the cross-section, to show
how the proton energy loss rate depends on energy.

\subsection{Numerical Modeling}
\subsubsection{Electron/positron Pair Production}

The pair-production spectra were calculated using the results
of the Monte Carlo code 
described in Section 2.1.
Protons of a specific energy
were allowed to interact with isotropic monoenergetic
target photons and the energies of the products were tabulated.
The photon target energies
used were $x_0=10^{-6}$, $10^{-4}$, $10^{-2}$ and 
1 (in units of electron rest mass). 
Proton energies ranged from $\gamma_{\rm p}=x_0^{-1}$ (so the
threshold requirement could be met) up to 
$\gamma_{\rm p}=10^4x_0^{-1}$ in logarithmic steps of 0.1. 

The pair-creation rate is then given
\eqb
\qbhinj(\gammae,t)&=&
\int d\gammap\,\nprot(\gammap,t)
\int dx\,
\nphot(x,t)\times
\nonumber\\
&&f(\gammap,\gammae,x)
\langle\sigma_{BH}(\gamma_{\rm p},x)\rangle
\eqe
where $f(\gammap,\gammae,x)$ is the distribution
found from the Monte-Carlo modelling 
(see eq.~\ref{pairdistr}), 
normalised such that
\eqb
\int d\gammae f(\gammap,\gammae,x)=2.
\eqe

As simple parameterisations of these curves      
do not give acceptable fits, we tabulated the
spectra and used interpolations
to derive the produced pair-injection spectrum
from a particular proton-photon collision.

We note that \citetalias{mastichiadiskirk95} used the approximation
$f(\gammap,\gammae,x)=2\delta(\gammae-\gammap)$ 
corresponding to $\xi=m_{\rm e}/m_{\rm p}$. For a proton
distribution $\nprot(\gammap)$ which decreases
as $\gammap$ increases, this overestimates
the production rate of electrons of a given energy.
 
\subsubsection{Proton losses}

Since the proton loses a small amount of energy 
(typically given by $\Delta\gamma_{\rm p}\sim\melec/\mprot$ 
in each pair-producing collision 
we can treat the losses  as a continuous
process and write 
\eqb
\protloss={{\partial}\over{\partial\gammap}}\left[\gammap n_{\rm p}(\gammap,t) 
Y(\gammap,t)\right]
\label{plosses}
\eqe
where 
\eqb
Y(\gammap,t)=\int dx \photdens(x,t)
\langle \sigma_{BH}(\gammap,x)\rangle
\xi(\gammap,x)
\eqe
is the convolution of the normalised
collision rate with the inelasticity $\xi$ given by 
eqn.~(\ref{inelast}).

\section{Proton injection and black-body photon field}
As a first example, we examine the case of relativistic proton
injection and subsequent cooling on a black-body photon field.
We assume, 
as usual in 
astrophysical cases, a power-law proton injection
of the form
\eqb
\protrat(\gamma)=\protin 
\gamma^{-\beta}H(\gamma-\gammapmin)H(\gammapmax-\gamma)
\label{protinj}
\eqe
where $\gammapmin$ and $\gammapmax$ are respectively
the lower and upper cutoff of the proton distribution. 
We also assume that the protons can 
escape at a rate $t_{\rm p,esc}^{-1}$ from the source region, taken 
to be a sphere of radius $R$. Then the stationary solution, 
in the absence of proton losses is simply 
\eqb
\protdens(\gamma)={\protin\tpesc}
\gamma^{-\beta}H(\gamma-\gammapmin)H(\gammapmax-\gamma).
\label{injss}
\eqe

We assume next that a significant number of 
external photons is
present and that these have a black-body
distribution of temperature $\tempbb$.
The energy density $U_{\rm BB}$
of these photons in the source can be parameterised 
in terms of the 
black-body compactness $\lbb$, defined by the relation
\eqb
\lbb={{U_{\rm BB}\sigmaT R}\over{m_{\rm e}c^2}}
\eqe
This emission can arise, for example, from the surface of 
an accretion disk located close to the source.
In general, if the source is irradiated by a black body at temperature 
$\tempbb$ whose surface occupies a solid angle of $\Delta\Omega$ as 
seen from the source, then
\eqb
U_{\rm bb}&=&\left({\Delta\Omega\over 4 \pi}\right)
a_{\rm rad}\tempbb^4
\eqe
where $a_{\rm rad}$ is the radiation density constant. In terms of 
compactness:
\eqb
\lbb&=& 615\times\left({\Delta\Omega\over 4 \pi}\right)
\left({\tempbb\over10^5\,\textrm{K}}\right)^4
\left({R\over10^{15}\,\textrm{cm}}\right)
\eqe 

We further assume that initially there are no other photons or
electrons present (except maybe in very small numbers)
and that there is no external electron injection.
Since the Bethe-Heitler pair production has a threshold condition
$\gamma x\ge 1+{\melec/\mprot}$ (corresponding to head-on collisions)
it follows that, in the case where
$y=\gammapmax(k\tempbb/\melec c^2)\ll 1$,  
there will be negligible pair
production. Therefore, for all practical purposes, the solution
of the proton equation will still be given by 
Eq.~(\ref{injss}). However, 
for $y\sim1$, the proton-photon
reaction rate is substantial and cannot be neglected
as a proton loss mechanism. Moreover, the produced 
electron/positron pairs provide  
an injection term for the electron equation and lose
energy mainly by inverse Compton scattering and/or synchrotron 
radiation.
As a result a photon spectrum is 
formed.
To investigate the properties of this emission, such as its 
luminosity and spectral shape etc., we must distinguish between two cases:

\subsection{Case 1: $\lprot\ll\lbb$ and negligible 
synchrotron losses}

In this case, the photons produced by electrons created in the BH 
process are not important as targets and the 
resulting photon spectrum is quite
simple. It is a single power-law of (number) spectral index $-3/2$,
extending up to an energy $\xmax$, that 
is approximately
the inverse
of the temperature of the black-body field, i.e.
$\xmax\simeq\Theta^{-1}$, where 
$\Theta=(k\tempbb/\melec c^2)$.  
The explanation of this spectrum 
is straight-forward: protons pair-produce on
the black-body field and, 
since they are produced with high energies,
the pairs cool on the black-body photons initially by
Compton scattering in the Klein-Nishina regime. This produces $\gamma$-rays
that are above the threshold for pair
production on the black-body photons and, therefore,  
are re-absorbed.  Lower energy pairs, which cool by Compton scattering in 
the Thomson regime, produce photons which are
below the threshold for photon-photon pair production. This 
naturally produces
an electron
distribution function $\nelec\propto\gamma^{-2}$
and thus a photon spectrum $\nphot\propto x^{-3/2}$.
The value of $\xmax$, therefore, is set by the condition
$\tau_{\gamma\gamma}(\xmax)\simeq 1$, 
where $\tau_{\gamma\gamma}$ is the optical depth
for photon-photon pair production on the background
black-body field.

\begin{figure}
\centering
\includegraphics[width=8.5cm]{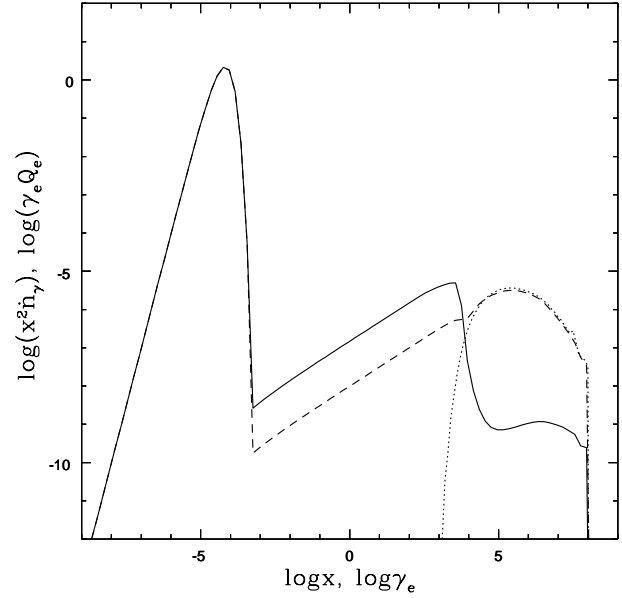}
\caption{Steady-state
photon spectrum resulting from a power-law
proton injection and subsequent proton-photon
pair production on a black-body photon field
in the case where photon-photon pair production (i)
has been ignored (dashed line) and (ii) has been
taken into account (full line). The dotted line curve
depicts the  electron/positron distribution function
at production.
For this particular run the protons were assumed to
be injected with a power-law of slope $\beta=2$,
between the limits
$\gammapmin=10^{0.1}$ and $\gammapmax=10^6$.
Also the values
$\lprot=2.5~10^{-3}$ and $\tpesc=\tcross$
have been assumed.
The black-body photon field parameters 
were $\tempbb=10^5\,$K ($\Theta=1.7~10^{-5}$) and $\lbb=1$. 
Synchrotron losses were neglected. 
\label{bbphspct}}
\end{figure}

Since for the present case there is
neither escape of electrons from the system
nor any sink of energy, (e.g. synchrotron self-absorption)
other than photon escape, 
the radiated  luminosity
(or, equivalently, $\lphot$) equals the luminosity injected
in pairs once a state steady is achieved, i.e.
$\lphot\simeq\lpairinj$, where 
\eqb
\lpairinj={1\over 3}\int d\gamma(\gamma-1)\qbhinj
\label{BHcompact}
\eqe
is the compactness of the created  pairs
from the Bethe-Heitler pair production.
Therefore, the overall photon luminosity depends
on the parameter $y$ defined above and, as we 
show below, also on $\lbb$
and $\lprot$.

As an illustrative case we take 
$\lbb=1\gg\lprot=2.5~10^{-3}$.
Here also $\tempbb=10^5$ K, while $\gammapmin=10^{0.1}$,
$\gammapmax=10^6$. The dotted line in
Fig.~\ref{bbphspct} depicts the electron injection function
which shows a broad maximum. The same
figure shows the photon
spectrum which is obtained from the cooling of
these electrons and the ambient black-body field.
The dashed line shows the
spectrum in the case where the $\gamma\gamma$
pair production has been artificially 
switched off.
The spectrum is flat and peaks at high energies. 
We note that both the electron production
function and the unabsorbed photon spectrum extend to
about two orders of magnitude above $\gamma_{p,mx}$,
an effect that is due to the kinematics of the Bethe-Heitler
pair production -- see Fig. 3.
At lower energies
it produces the characteristic $-3/2$ power-law.
The full line shows the
final photon spectrum which includes $\gamma\gamma$
absorption. It has still the same power-law,
however all the details of the electron injection
spectrum which were evident in the unabsorbed spectrum
have disappeared due to the intense attenuation.
Nevertheless the overall luminosity is conserved (as it
should be) and as the electromagnetic cascade
redistributes the power to lower energies,
the flux is increased there. 
Figure 4 shows the effects of losses on the 
steady-state proton distribution for these parameters ($\ell_{\rm BB}=1$) and two 
higher values of $\ell_{\rm BB}$, compared to the loss-free case.

\begin{figure}
\centering
\includegraphics[width=8.5cm]{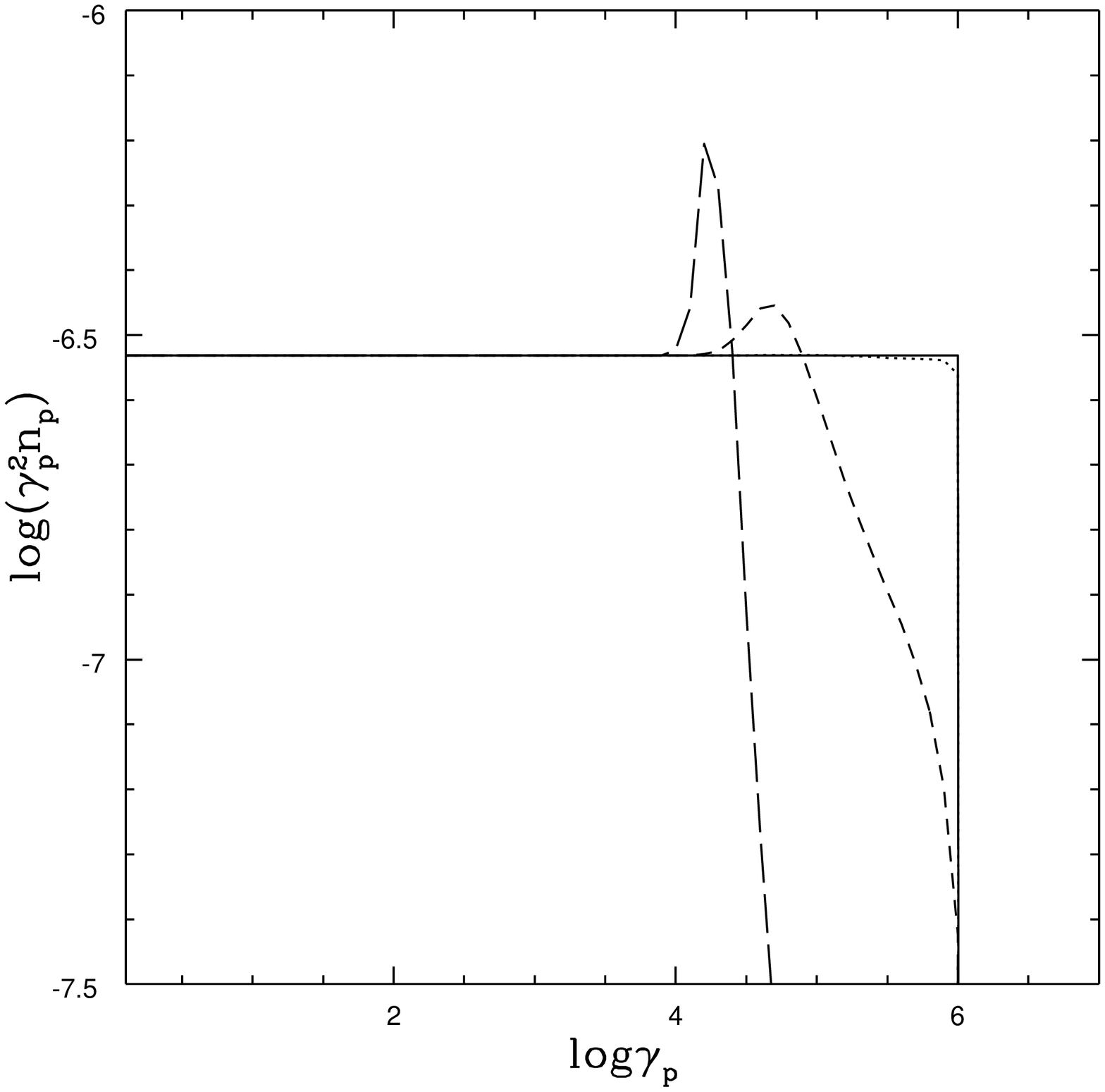}
\caption{Proton steady-state
spectra for various external black-body 
photon compactnesses. All cases
are taken for the same injection  
proton parameters ($\lprot=2.5~10^{-3},
~\gammapmax=10^6,~\tpesc=1$)
and for the same temperature of the external
black-body field ($\tempbb=10^5~K$)
The plotted proton spectra are shown 
when $\lbb=0$  (no-loss case, full line), 
$\lbb=1$ (dotted line), $\lbb=10^2$ (short-dashed line)
and $\lbb=10^4$ (long-dashed line).
\label{bbprspct}}
\end{figure}

The efficiency of the BH process, i.e., the ratio of the proton 
power turned into pairs to the total power injected as protons, is 
shown in Fig.~5 as a function of the maximum Lorentz factor of 
injection $\gamma_{\rm p,max}$. The inelasticity of this process is small 
($\xi\sim10^{-3}$), so that high efficiency can only be achieved if 
a proton interacts many times before escape. This is indeed the case 
for high black-body compactnesses --- the efficiency exceeds $40\%$ 
for $\ell_{\rm BB}>10^4$ and $\gamma_{\rm p,max}>10^7$.

\begin{figure}
\centering
\includegraphics[width=8.5cm]{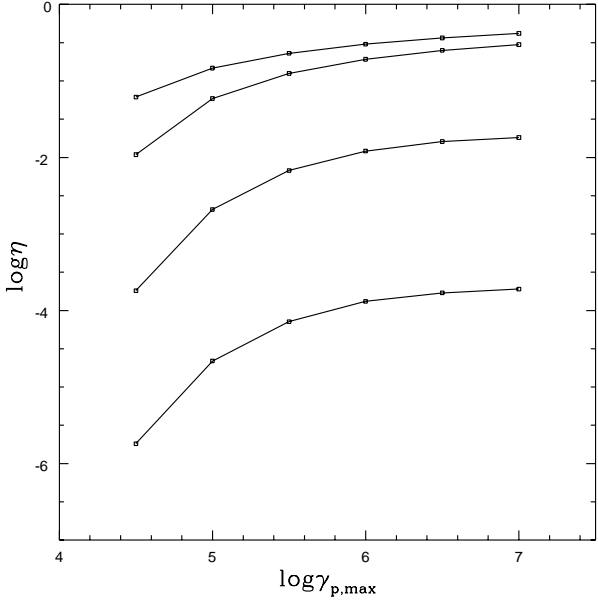}
\caption{Proton efficiency, i.e.
fraction of power lost to pairs 
to total power injected in protons
as a function of the upper cutoff 
of the proton distribution $\gammapmax$.
The rest of the  proton injection  
parameters are kept constant 
($\lprot=2.5~10^{-3}$, $\gammapmin=10^{0.1}$, 
$\tpesc=1$),
while the external
black-body field has temperature $\tempbb=10^5~K$.
and compactnesses  (bottom to top)
$\lbb=.01,~1,~100,~10^4$. 
\label{bbeffic}}
\end{figure}

\subsection{Case 2: $\lprot\ll\lbb$ and non-negligible
synchrotron losses}

\begin{figure}
\centering
\includegraphics[width=8.5cm]{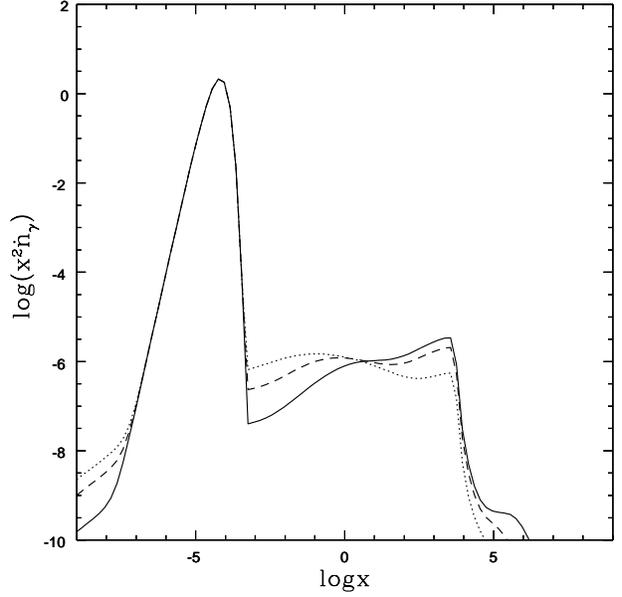}
\caption{
Steady state photon spectra for
the same parameters used in Fig.~\ref{bbphspct}
but including synchrotron radiation. The magnetic
field used was $B=100~G$ and the magnetic compactnesses
were $\lb=0.01$ (full line), $\lb=0.1$ (dashed line)
and $\lb=1$. The black-body compactness
was $\lbb=1$. 
\label{bbsyn}}
\end{figure}

In the more realistic case, where
synchrotron losses cannot be neglected, the
spectrum of the electrons becomes more complicated,
as now the electrons cool by a combination of 
synchrotron radiation and inverse Compton scattering.
Fig.~\ref{bbsyn} shows the obtained spectra
in the cases where the magnetic compactness $\lb$, defined according
to 
\eqb
\lb&=&\sigmaT R \left( {B^2\over {8\pi \melec c^2}}\right)
\eqe
is comparable to $\lbb$.
The redistribution of the radiated luminosity
to increasingly lower energies is evident, the reason
being that, for this particular choice of
$\tempbb$ and $B$, inverse Compton scattering
produces much harder photons than the synchrotron
mechanism. Therefore, as $\lb$ progressively increases, 
the peak is shifted towards the lower energies, showing
the increasing importance of synchrotron radiation
as an electron energy loss/photon emission mechanism.
As in Fig.~\ref{bbphspct},
the spectrum continues to be strongly 
absorbed for photon energies 
$x>\xmax\simeq\Theta^{-1}$. 

\section{The Pair-Production/Synchrotron Instability}

\subsection{Marginal stability}

\citet[henceforth KM92]{kirkmastichiadis92}
\defcitealias{kirkmastichiadis92}{KM92} 
have shown that ultra-relativistic protons
can, under certain conditions, become unstable
to various types of radiative instability. They showed 
explicitly the necessary conditions for one of them to
occur, namely the Pair-Production/Synchrotron instability
(henceforth PPS). 
To understand the basic idea, assume that protons are
confined in a region of characteristic radius R  
where a magnetic field of strength $B$ is also present.
Assume, moreover, that the protons are relativistic and
have Lorentz factors such that if they photo-pair produce,
the synchrotron photons radiated from the created pairs
are 
sufficiently energetic for the protons  
to produce more pairs on them. 
Making the simplifying
assumptions that (i) the created pairs have the same
Lorentz factors as the protons and (ii) the synchrotron
photons are all emitted at the critical frequency, 
\citetalias{kirkmastichiadis92} showed that in order for protons to be able to
initiate this loop they should have
Lorentz factors above a critical value given by 
\eqb
\gcrit=\left({{2}\over {b}}\right)^{1/3} 
\eqe
where $b=B/\bcrit$ with $\bcrit$ the critical value
of the magnetic field ($\bcrit=\melec^2 c^3/e\hbar=
4.413~10^{13}~$Gauss).
For this loop to be self-sustained it is  necessary that
at least one of the synchrotron photons should produce
a pair before escaping the source and this condition
naturally leads to a critical proton number density
\citepalias[see][eq.~6]{kirkmastichiadis92}. 

\citetalias{mastichiadiskirk95} 
presented a numerical simulation of the PPS instability
in the case where protons are accelerated from low momenta
by a Fermi-type acceleration scheme. Once the protons
(assumed to have a density exceeding the critical number density)
reached energies above $\gcrit$, the 
conditions for the instability loop were complete and the internally 
produced photons increased, saturating the acceleration, and
driving the system to equilibrium.
However, the code used by \citet{mastichiadiskirk95} is limited by
the 
simplifying assumptions mentioned in the previous paragraph.
Here we re-examine 
this problem with the 
improved version of the code that uses, as 
described
in section~3, the Bethe-Heitler pair-production spectra as given by
Monte-Carlo code and the full synchrotron emissivity. The objective
is to find an accurate estimate of the critical
proton number density above which the PPS instability occurs, 
i.e., a numerical version of the analytical (but approximate) 
Fig.~1 of \citetalias{kirkmastichiadis92}.

To make our results directly comparable with those of the aforementioned
figure, we have set the parameters to the values prescribed there.
For this we took a source size $R=10^{15}$ cm,
a magnetic field $B=10^3$ Gauss and
a proton distribution function of the form $\nprot(\gammap)=
\nprin\gammap^{-\beta}$ with $\beta=2$
from the lowest allowed proton energy $\gammapmin=10^{0.1}$
to a maximum energy $\gammapmax$. We note that
in this case the proton distribution is held constant throughout each run,
i.e. protons do not evolve.
For various values of $\gammapmax$,
we run the code for different 
values of the only remaining free parameter ($\nprin$). 
According to \citetalias{kirkmastichiadis92}, the time evolution 
of the photon and electron distribution
functions is of the form $\nphot(t)\propto\nelec\propto e^{st}$
with $s>0$ when the protons are in the unstable regime.
Thus in order to verify numerically the existence of the
PPS instability we seek  
a value $\nprincrit$ above which the internally produced
electron/positron pairs and photons start increasing with time.

\begin{figure}
\centering
\includegraphics[width=8.5cm]{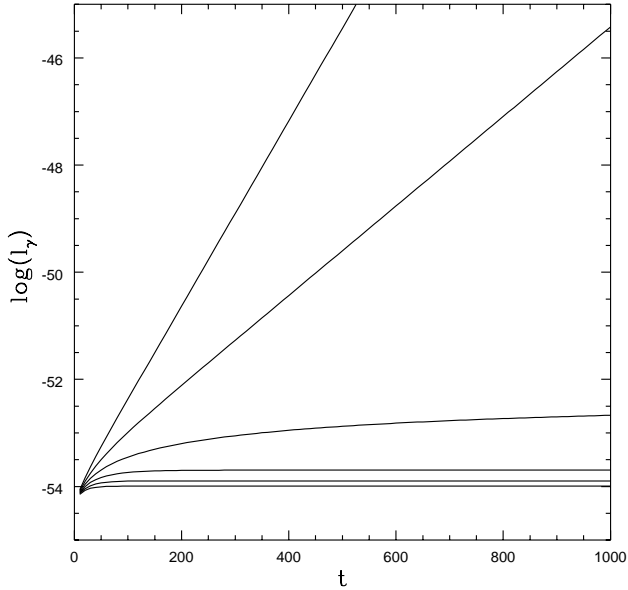}
\caption{
Plot of the internally produced photon compactness $\lphot$
as a function of time for values of $\nprin$ around
the critical value $\nprincrit$. In each curve the value 
of $\nprin$ is increased 2\% from its previous value.  
For these runs $\gammapmax=10^6$.
\label{instgrow}}
\end{figure}

The onset of instability 
for $\gammapmax=10^6$ can be seen
in Fig.~\ref{instgrow} which 
depicts the photon compactness $l_{\gamma}$
as a function of time $t$ (expressed, as always, in units of $\tcross$)
for various values of $\nprin$ around  $\nprincrit$. 
When the protons are in the
stable regime there is some pair production between
the protons and the synchrotron photons produced
from the initial electron distribution but
the system, for times larger than  
the synchrotron cooling time,
settles to a steady-state. Thus for $t>\tcool$ 
we get $s=0$.
However, as can be seen from the figure,
as $\nprin$ increases (from bottom to top)
the photons start to grow exponentially,
with $s$ increasing with increasing $\nprin$.
It is worth
mentioning that each curve corresponds to a value
of $\nprin$ that is 
larger by only 2\% than its previous value;
therefore, this figure depicts the 
rapid onset of the instability.  

\begin{figure}
\centering
\includegraphics[width=8.5cm]{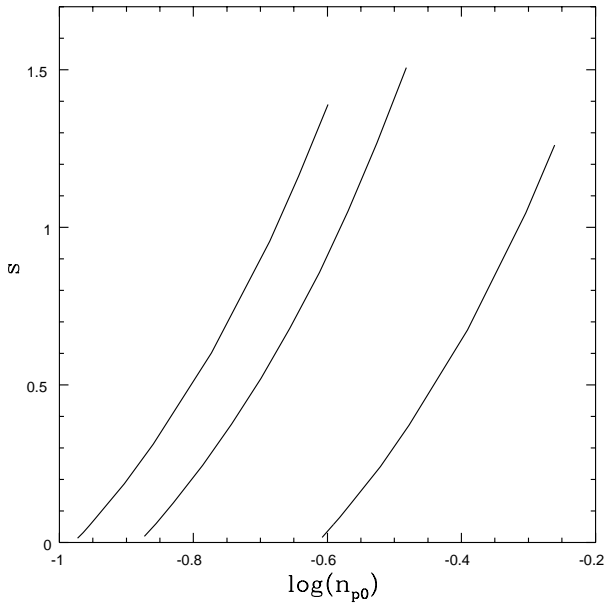}
\caption{
Behaviour of the instability growth index $s$ as a
function of $\nprin$ for values of 
$\gammapmax=10^5,~10^6,~10^7$ (right to left).
\label{splot}}
\end{figure}

Fig.~\ref{splot} shows the behaviour of $s$ as a function
of the proton normalisation $\nprin$ for three
values of $\gammapmax$. It is evident
that once the instability sets in, $s$ is a very sharp function
of $\nprin$. Therefore, we find that increasing $\nprin$ 
approximately by a factor
of 2 above its critical value, $s$ becomes
greater than one, i.e. the density of photons starts growing
on a  timescale shorter than the crossing time of the source.
This, as we shall see in the next section, has catastrophic
consequences for the high energy protons as the 
spontaneously growing
photons make them lose their energies.

\begin{figure}
\centering
\includegraphics[width=8.5cm]{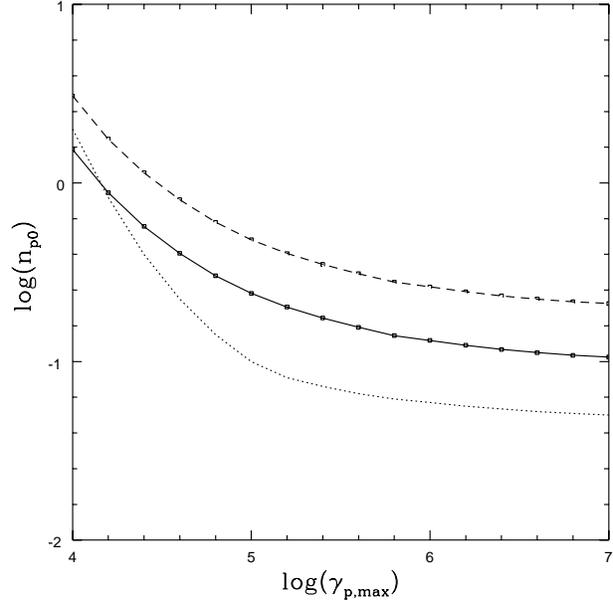}
\caption{
Plot of the numerically obtained
marginal stability as a function of $\gammapmax$ (full line).
The dashed line shows the $s=1$ locus, while the dotted line
depicts the marginal stability results of \citetalias{kirkmastichiadis92}. 
Note that the \citetalias{kirkmastichiadis92}
approximation is reasonably good at low $\gammapmax$.
\label{naturev}}
\end{figure}

Fig.~\ref{naturev} shows the marginal stability curve
as is obtained from the present code (in practice we have
calculated the values of $\nprin$ which correspond to $s=.05$.
Due to the steep dependence of $s$ on $\nprin$ this value
can be considered as  very close to
the  marginal stability one). Note
that this curve is in very good agreement 
with
the
curve estimated by 
\citetalias{kirkmastichiadis92} (plotted here as a dotted line)
for values of $\gamma$
close to the threshold, but exceeds it by a factor of about 2 for
at high energy.
This difference can be attributed to the overestimation 
of the electron production rate in the
BH process as discussed in Section 2. For maximum values of the proton Lorentz 
factor close to $\gamma_{\rm crit}$, all the BH interactions occur 
close to threshold, so that the assumption used by MK is 
accurate. 
(The small difference in the shape of the KM curve and our
present result in this energy range can be attributed 
partly to the kinematics
of Bethe-Heitler pair production and partly to the fact that
the full expression for the synchrotron emissivity was 
used).
However, once the upper cut-off of the proton distribution 
substantially exceeds $\gamma_{\rm crit}$, the effect of pairs 
injected with $\gamma>\gamma_{\rm p}$ becomes important. These 
are not taken into account in the approximation used by MK92, 
but are treated accurately in the simulation-based method used here. 
It is evident that the basic concept of the feedback loop 
remains unaltered by our more accurate treatment. The 
quantitative implications, however, are discussed below.

\subsection{Proton injection}

In order to see  the effects of the PPS instability 
when proton losses are taken into account
we assume once again (see section 4) that
protons are injected in a region of radius $R$ with 
a power-law (eq.~\ref{protinj}). However, to make the picture
less complicated, we assume that there is no external
photon field present. Thus one expects that the proton
spectrum will reach an equilibrium state given by 
the no-loss solution (eq.~\ref{injss}). This is indeed true as
long as this steady-state solution is below the critical
density for the PPS instability, i.e. when
$\nprin={\protin\tpesc}<\nprincrit(\gammapmax)$,
with $\nprin$ the normalisation of the protons and
$\nprincrit(\gammapmax)$ given from Fig.~\ref{naturev}.
In this case also there
is a very small number of internally produced photons.
However, as soon as the condition
$\nprin={\protin\tpesc}>\nprincrit(\gammapmax)$ holds,
the criteria for the PPS instability are satisfied.
As a result,
the photons increase exponentially, protons lose 
energy due to pair production and the system 
behaviour depends on the choice of the parameters
$\protin$ and $\tpesc$. 

\begin{figure}
\centering
\includegraphics[width=8.5cm]{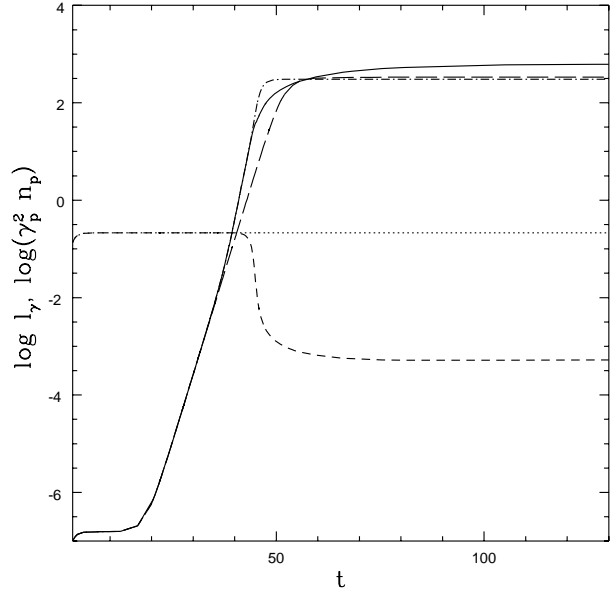}
\caption{
Plot of the evolution of the system 
in the case where protons are injected with
$\beta=2,~\gammapmax=10^6,~\tpesc=1$,
while the combination of the 
proton injection rate ($\protin=0.22$)
and $\tpesc$ is such that the steady-state
proton distribution in the no-loss case
corresponds to an unstable proton distribution.
The solid line shows the photon compactness
when all processes are taken into account,
while the long dashed and dot-dashed lines
show the photon compactness when certain processes
are omitted (for details see text). 
The dotted and
short dashed lines show the evolution of the
first and last proton occupation number bin
in the case where all the relevant processes are taken into 
account.  
\label{steady1}}
\end{figure}

As a first example, we show in Fig.~\ref{steady1}
the case where 
$\beta=2,~\gammapmax=10^6,~\tpesc=1$
with a proton injection rate of $\protin=0.22$.
The proton compactness parameter, according to eq.~\ref{procompact},
is $\lprot=1710$.
The above combination of $\protin$ and $\tpesc$,
according to Fig.~\ref{splot}, corresponds to a
feedback loop that causes 
the photon density to increase with $s=0.7$. 
The dotted and short-dashed lines
show, respectively, the evolution of the lowest and highest 
differential density bins
of the proton energy. 
In agreement with the analytical solution of  
Eq.~(\ref{protkinet}) in the no-loss case, the number of particles in these
bins increases
very quickly (in about one $\tcross$) to a steady state, 
which, 
however, corresponds to an unstable proton configuration.
The long-dashed line shows the photon compactness
when only synchrotron radiation is taken into account.
This increases as $\lphot\propto e^{st}$ with $s=0.7$ until
it reaches a steady state. The dot-dashed line shows
the effects on the photon compactness caused from the addition
of inverse Compton scattering and photon-photon pair production.
As these processes are quadratic with respect to the photon
and electron/positron number densities they do not affect
the slope of the compactness at early stages, i.e. as long as
$\lphot\ll 1$. However, at the later stages of evolution
these processes become important and, because they are quadratic,
they cause the photon compactness to increase even faster
and reach saturation sooner. Finally the solid line shows 
the evolution of the photon compactness when, in addition
to the above processes, photon trapping due to 
the high 
density of 
created pairs
is taken into account. This leads to higher photon
compactnesses as relatively more energy is extracted from 
the high energy protons. This curve must be considered
the `correct' one as it contains all the relevant processes.
On the other hand the photon increase 
causes the high energy protons (short-dashed line)
to lose energy and settle
in a new steady state. These losses do not affect naturally 
the low energy protons (dotted line)
which 
maintain their original steady state. 

\begin{figure}
\centering
\includegraphics[width=8.5cm]{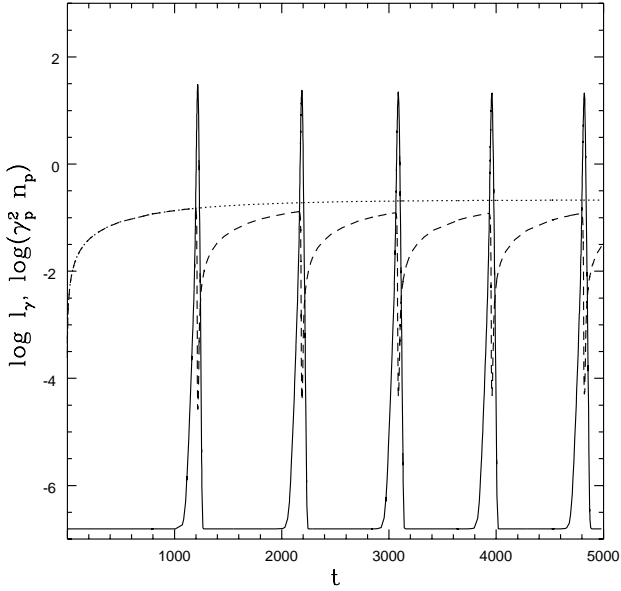}
\caption{
Plot of the evolution of the system 
in the case where a low
proton injection rate 
($\protin=2.2~10^{-4}$) combines with a slow
proton escape time
($\tpesc=10^3$) in such a way
that the steady-state
proton distribution in the no-loss case to
correspond to an unstable proton distribution.
The rest of the parameters are as in the previous 
figure. Solid line  shows $\lphot$, while
dotted and
short dashed lines show the evolution of the
proton spectrum at $\gammap=10.^{0.1}$ and
$\gammap=10^6$ respectively.
\label{oscill}}
\end{figure}

At the other extreme, one can envisage a case where protons 
are 
injected slowly, but have a very long  
escape time. 
In this case we take
$\protin=.22~10^{-3}$ and $\tpesc=10^3$ which corresponds
to the same no-loss steady state as before.
The result is shown in Fig.~\ref{oscill}, where
quasi-periodic behaviour typical of a relaxation 
oscillator is apparent.
The protons accumulate in the source
and, once
their density 
rises
above the critical density,
the photon 
density grows rapidly 
(on the timescale of a few times
$\tcross$) and deplete the high energy 
part of the proton spectrum. Once the protons have
lost their energy, there is nothing left to sustain the 
loop and the photons escape from the system. This cycle is
repeated 
as the protons accumulate again
in the source.
The behaviour of the lowest and highest proton energy bins
is also shown (dotted and dashed lines respectively). 
It is clear that photons and high energy protons are
anticorrelated, in the sense that when one population is
high, the other is low. These cycles are similar to those
found by \citet{sternsvensson91} using Monte-Carlo techniques.

\section {Simultaneous electron and proton injection}

As a last case we examine the situation where electrons and protons are 
injected simultaneously, i.e., both proton and electron kinetic
equations~(\ref{protkinet}) and (\ref{eleckinet}) have an external
injection term. We assume that both of these terms are in power-law form,
so that,
in addition to Eqn.~(\ref{protinj}) that
describes proton injection, we prescribe electron injection
using a similar expression: 
\eqb
\elecrat(\gammae)=\elecin 
\gamma^{-\beta}H(\gammae-\gammaemin)H(\gammaemax-\gammae).
\label{elecinj}
\enspace.
\eqe

To lower the number of free parameters we assume that both electrons 
and protons are
injected with the same spectral index $\beta$ and with the same
maximum Lorentz factor, i.e. we set arbitrarily
$\gammaemax=\gammapmax$. 
Moreover, when solving the relevant kinetic equations, we assume that 
the two species have equal escape times, i.e., $\tpesc=\teesc$. 

Depending on the particular choice of parameters,
high energy electron injection can result in a
synchrotron and/or an inverse Compton component 
\citep[see, for 
example,][]{mastichiadiskirk97}.
When protons are injected as well, these will interact
with the aforementioned photons 
causing a secondary injection of Bethe-Heitler pairs.
This leads to a non-linear situation. 
To see this, one
should compare the compactness of the externally injected 
electrons  $\lelecext$ (eq.~\ref{extelinj}) with 
the corresponding compactness of the internally produced pairs
via the Bethe-Heitler process $\lpairinj$ (Eq.~\ref{BHcompact}). 
As the latter is, in general, a function of both $\lprot$
and $\lelecext$, when $\lelecext < \lpairinj$ 
the system operates in the non-linear regime, in the sense
that the cooling of the protons occurs mainly on the internally
produced photons.

\begin{figure}
\centering
\includegraphics[width=8.5cm]{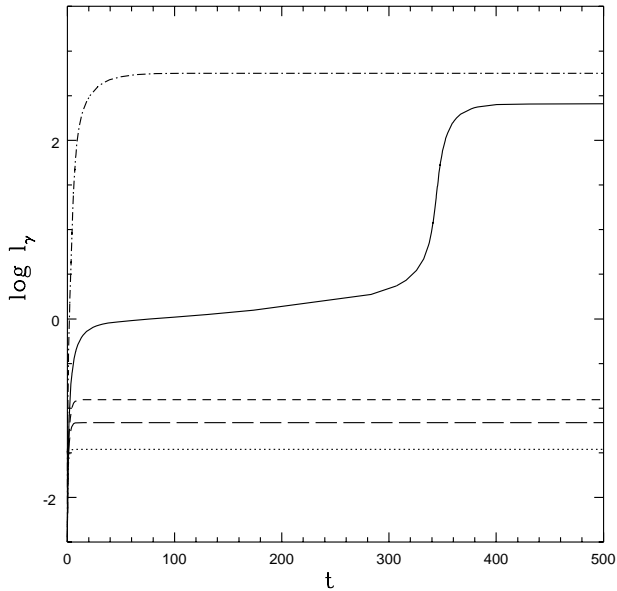}
\caption{
Plot of the evolution of the photons 
in the case where protons and electrons are
injected simultaneously with the same slope 
$\beta=2$ and the same high Lorentz factor cutoff,
i.e. $\gammapmax=\gammaemax=10^6$ in a region of radius
$R=10^{15}\,$cm immersed in a magnetic field of strength
$B=10^3\,$G.
The injected electron compactness is $\lelecext=0.04$ and the
evolution of the photons is shown 
for
proton compactnesses 
(bottom to top) $\lprot=0$, $200$, $400$, $800$ and $1600$.
Both species have an escape time  equal to $\tcross$. 
All except the highest curve
correspond to stable proton distributions
in the no-loss case. Nevertheless the 
solid curve shows 
a loop 
that
eventually leads to catastrophic proton losses.
The uppermost curve (dot-dashed line) 
corresponds to a 
proton distribution that is unstable even
in the no-loss case.
Therefore the photons
increase quickly and drive the system to an equilibrium.   
\label{extellc}}
\end{figure}

To investigate the effect described above we 
proceed as follows: 
we keep the electron injection parameters constant and change
only the normalisation of the proton injection rate $\protin$ 
in eq~(\ref{protinj})
--- or, equivalently, we change
the proton compactness parameter $\lprot$ 
(Eq.~\ref{procompact}).
When $\lprot\ll\lelecext$ the resulting photon spectrum is simply 
that produced from the cooling of the externally injected electrons.
However as we increase $\lprot$, the quantity $\lpairinj$ 
increases as well and, at some stage, it becomes comparable to $\lelecext$.
Above this point the system enters the nonlinear regime, as the cooling
of the protons occurs primarily on its own radiation. Finally, 
above some critical
proton compactness, a loop analogous to the one described in Section~5 
operates, and the protons convert a substantial fraction of their energy
content to electron/positron pairs and radiation.      
This is depicted in Fig.~\ref{extellc} which shows the evolution of the
photons when $\lelecext$ is kept constant, while $\lprot$ varies.
The first curve (dotted line) assumes that no protons are injected.
Due to the particular choice of the parameters the primary electrons
cool fast and the system 
quickly reaches 
a steady state.
In the next two cases, the injected proton compactness is 
$\lprot=200$
(long-dashed line) and $\lprot=400$ (short-dashed line) respectively. Again
a steady state is quickly reached. However the photon compactness $\lphot$
increases as $\lprot$ increases because of the BH pairs created
and subsequently cooled.
In both of these cases the feedback loop 
does
not operate in the sense that the proton losses 
remain
low --- or, equivalently,
$\lphot\simeq\lelecext+\lpairinj\ll\lprot$.  
However, for
the last two cases which are for $\lprot=800$ (full line) and $\lprot=1600$
(dot-dash line) the effects of the feedback are evident.
In the case of the 
solid curve, and
for the first 300 or so light crossing times, there is a gradual increase 
of pairs and photons in the system until their numbers are built to a level
that allows a catastrophic release of the energy stored in protons. 
After that, a steady state is reached, but at a level much 
higher than that of the
previous cases: $\lphot\simeq\lelecext+\lpairinj\simeq\lprot$.   
It is worth mentioning that this loop 
operates at a proton compactness which is
below the critical threshold obtained in Section~5. Therefore, the presence of
external electrons helps to initiate the catastrophic proton energy losses
at lower proton densities.
Finally, the uppermost curve corresponds to a proton injection which is above
the critical threshold for the PPS instability and the photons grow very
quickly as discussed in Section~5 --- see also Fig.~\ref{steady1}.

\begin{figure}
\centering
\includegraphics[width=8.5cm]{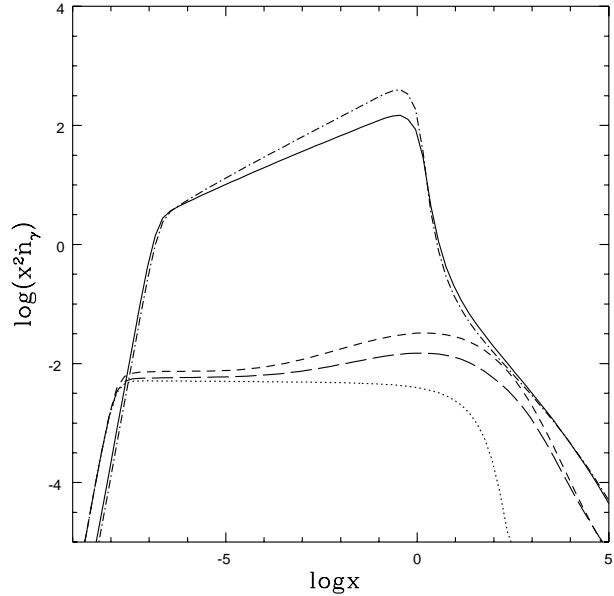}
\caption{
Plot of the photon spectra corresponding to the
steady states obtained for the runs shown in
Fig.~\ref{extellc}. The bottom curve (dotted)
shows the spectrum obtained from the cooling of the
injected electrons when no protons are injected.
As the injected proton compactness increases the
photon spectrum is modified due to the presence of
the radiation from the created pairs (long and short-dashed curves). 
Finally when the loop operates
the photon spectrum is strongly absorbed above 1~MeV by
photon-photon pair creation (solid and dot-dashed curves).    
\label{extelsp}}
\end{figure}

The photon spectra corresponding to the steady states obtained
in each of these runs are shown in Fig.~\ref{extelsp}. The bottom curve 
(dotted line) corresponds to 
the injection of  
electrons only 
whereas
the four others are for electron and proton injection, 
corresponding 
to the curves of Fig.~\ref{extellc}.
The extra component
due to BH pair production is apparent on the two lower spectra that include
protons (long and short-dashed lines).  
The two uppermost curves
correspond to the steady-state spectra when the loop 
was able to extract most of
the proton energy. Due to the resulting high photon 
compactness the spectra 
are strongly absorbed at energies above 1~MeV due to photon-photon pair production.

\begin{figure}
\centering
\includegraphics[width=8.5cm]{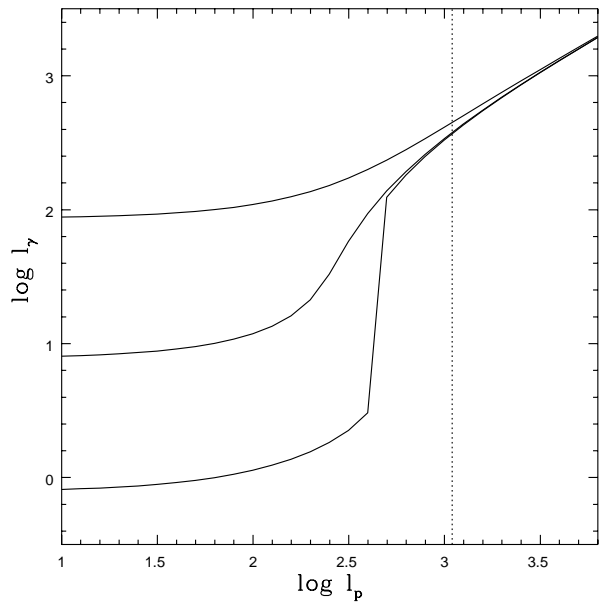}
\caption{
Plot of the
photon compactness versus the proton compactness
for various injected electron compactnesses.
The curves correspond to $\lelecext=1$, 
$10$ and $100$ (from bottom to top).
The other parameters are the same as those used
in Figs.~\ref{extellc} and \ref{extelsp}.
The dotted line represents the critical proton
compactness in the no-loss case as this was
described in Section~5.
\label{plotlg}}
\end{figure}

The above findings
are summarised in Fig~\ref{plotlg} which shows the photon 
compactness of the system versus the proton compactness for various
injected electron compactnesses.  For low values of $\lprot$
(i.e. less than 30), the photons of the low-frequency part of the SED
produced come almost  exclusively from
the presence of the electrons, i.e. the protons cannot significantly
influence the low-frequency behaviour of
the system.  We note here that for synchrotron proton blazar model
fits to the SED of BL Lac Objects observed at gamma-ray energies
(e.g. \citep{mueckeetal03}) the proton compactness is $\lprot \sim
10^{-3}$--$10^{-2}$ implying that Bethe-Heitler pair production, and
any associated instability, is unimportant in these models.  
 For intermediate values of $\lprot$ (i.e. between 30 and
300) the internally produced Bethe-Heitler pairs make their presence
visible in that their cooling increasingly dominates the photon
spectrum.  Finally at even higher compactnesses the protons become  
unstable and cool efficiently on their 
\lq\lq own\rq\rq\ radiation.

\section{Summary/Discussion}

In the present paper we have examined some
consequences arising from the presence of
ultrarelativistic hadrons in compact sources. This was done with the help
of a  numerical code that was constructed to follow the evolution
of the system through the solution of three coupled, time-dependent kinetic
equations for protons, electrons and photons respectively.
All the 
relevant
basic processes 
involving electrons and photons in astrophysical 
pair-plasmas were included. 
The coupling between the hadronic and the leptonic component was assumed to
occur via Bethe-Heitler pair-production. For this process, detailed 
electron/positron pair-production spectra were obtained with the help of
a Monte-Carlo code. These were  then  incorporated into the kinetic
equations 
which were subsequently solved numerically, revealing effects 
mainly due to synchrotron
and inverse Compton losses. 

The choice of the kinetic equation approach allowed us to study various 
aspects of the behaviour of such a system. 
Thus we showed that the presence of an external black-body 
radiation field can 
extract 
energy efficiently from the relativistic protons only
when the photon compactness is high.

Of special interest are
the non-linear cases, i.e., cases in which the protons cool 
not on a 
prescribed 
external photon field, but
on the radiation of the internally produced
Bethe-Heitler pairs. In the present paper we have verified the
existence of the
'Pair-Production/Synchrotron' loop 
previously studied
analytically.
We showed that this is a very efficient
way of 
channelling
proton energy into electron/positron pairs and
radiation. 
and that the coexistence of
relativistic electrons in the system does not 
stabilise the system but, on the contrary,
lowers the critical density threshold, i.e. it facilitates the
efficient transfer of energy from the hadronic component to
the leptonic/photonic one.
This, and the fact that the threshold and critical density   
conditions can be greatly relaxed if the protons are in relativistic
bulk motion, makes this loop a promising candidate for some         
AGNs \citep{kazanasmastichiadis99} and GRBs \citep{kazanasetal02}.

Of the various processes omitted in the present treatment, 
the most important is that of 
photo pion-production. This, however, does not affect the results of the
present paper 
because,
in the examples shown, the initial conditions
were 
chosen so as to avoid the onset of this process.
Other processes involving protons, 
such as
proton synchrotron radiation and
proton-proton interactions, etc., 
are
negligible for the 
parameters of the particular examples given in this paper.

{\sl Acknowledgments:} A.M. would like to thank R.J.P. for hospitality
during his stay in the University of Adelaide. This research was funded in
part by a  Grant from the 
Special Funds for Research of the University of Athens.
A.M and J.K. acknowledge the EC funding under contract 
HPRCN-CT-2002-00321 (ENIGMA network).
The research of RJP is supported by an 
Australian Research Council Discovery Grant.


\end{document}